\documentclass{PoS}
\usepackage{braket}
\usepackage{bm}
\usepackage{color}

\newcommand{\ta}{\mathtt{a}}

\title{Matrix product states for Hamiltonian lattice gauge theories}

\ShortTitle{MPS for gauge theories}

\author{\speaker{B. Buyens, K. Van Acoleyen}$^a$, J. Haegeman$^{a}$, F. Verstraete$^{ab}$\\
\llap{$^a$}Department of Physics and Astronomy, Ghent University, Krijgslaan 281, S9, 9000 Gent, Belgium \\
\llap{$^b$}Vienna Center for Quantum Science and Technology, Faculty of Physics,
University of Vienna, Boltzmanngasse 5, 1090 Vienna, Austria
\\ E-mail: \email{boye.buyens@ugent.be}, \email{karel.vanacoleyen@ugent.be}, \email{jutho.haegeman@ugent.be}, \email{frank.verstaete@ugent.be}}

\abstract{Over the last decade tensor network states (TNS) have emerged as a powerful tool for the study of quantum many body systems. The matrix product states (MPS) are one particular case of TNS and are used for the simulation of 1+1 dimensional systems. In \cite{Buyens} we considered the MPS formalism for the simulation of the Hamiltonian lattice gauge formulation of 1+1 dimensional one flavor quantum electrodynamics, also known as the massive Schwinger model. We deduced the ground state and lowest lying excitations. Furthermore, we performed a full quantum real-time simulation for a quench with a uniform background electric field. In this proceeding we continue our work on the Schwinger model. We demonstrate the advantage of working with gauge invariant MPS by comparing with MPS simulations on the full Hilbert space, that includes numerous non-physical gauge variant states. Furthermore, we compute the chiral condensate and recover the predicted UV-divergent behavior.}

\FullConference{The 32nd International Symposium on Lattice Field Theory,\\
                23-28 June, 2014\\
                Columbia University New York, NY}

\begin{document}
\noindent\textit{\bf Introduction. }The dimension of the Hilbert space on a lattice with $N$ sites grows exponentially with $N$, which at present makes it impossible to diagonalize Hamiltonians exactly for $N \gtrsim 40$. But often we are interested in the low-energy states of a system, and it turns out that the area law for entanglement entropy \cite{AreaLaw} gives a universal identification of the physically relevant tiny corner of Hilbert space for these states. This is where TNS \cite{TNS} come into play: they are ans\"{a}tze that efficiently represent general low-energy states, by encoding the wave function into a set of tensors whose interconnections capture the proper entanglement behavior. As such the Hamiltonian TNS framework effectively replaces the linear Schrodinger equations on the full gigantic Hilbert space, with new non-linear equations on an exponentially smaller manifold. This has opened up the possibility of quantum many body simulations of real-time non-equilibrium dynamics and large fermionic densities, the two regimes that are inaccessible to Monte-Carlo simulations.
\\ \indent In one spatial dimension the MPS formalism, underlying the density matrix normalization group \cite{DMRG}, is the state of the art method for computing both static and dynamical properties of condensed matter systems. A MPS \cite{FannesNachtergaele} associates with each site $n$ of the lattice a tensor $A_{n}^q \in \mathbb{C}^{D_n \times D_{n+1}}$, where $\{ \ket{q}_n$ : $q = 1,\ldots, d\}$ spans the local $d-$dimensional Hilberts space of site $n$. Each tensor $A_{n}$ is connected with the tensors on the neighboring sites by matrix multiplication, the wavefunction of a MPS thus reads: $\ket{\Psi[A_1,\ldots,A_n]} = \sum_{\{q_n\} = 1}^d \mbox{tr}(A_1^{q_1}\ldots A_N^{q_n}) \ket{q_1}_1\ldots \ket{q_N}_N.$ $D_n$ are the virtual dimensions and $D = \max_n D_n$ is called the bond dimension of the MPS. For a local observable $O = \sum_n o_n$ where $o_n$ has only support on a small number of sites, the expectation value $\bra{\Psi[\bar{A}_1,\ldots,\bar{A}_n]}O\ket{\Psi[A_1,\ldots,A_n]}$ can be computed with complexity $\mathcal{O}(ND^5)$ \cite{VerstraeteMurg}. When $D_0 = D_{N+1} = 1$, we say that the MPS has open boundary conditions (OBC) and the computation time even reduces to $\mathcal{O}(ND^3)$. Also, for gapped nearest-neighbor Hamiltonians, it has been proven that one can efficiently approach the ground state with a MPS, i.e. a MPS where $D$ scales polynomially in $N$ \cite{MPSeffGS}. In the thermodynamic limit ($N \rightarrow \infty$), for translation invariant systems, one takes the tensors $A$ site-independent ($A_n^q = A^q \in \mathbb{C}^D, \forall n$) and one obtains a uniform MPS $\ket{\Psi(A)}$. Again, the ground state of a nearest-neighbor Hamiltonian can be approximated efficiently and all expectations values of local operators can be computed in $\mathcal{O}(D^3)$ time \cite{HaegemanMPSB}, allowing MPS simulations with bond dimension of order $1000$ on an ordinary desktop.
\\ \indent It has been realized recently that the TNS framework is also very well suited for the Hamiltonian simulation of gauge theories. For $d=1+1$ different groups have considered MPS simulations, mainly of the Schwinger model \cite{Buyens, MPSSchwinger}. For higher dimensions different gauge invariant TNS constructions have also been developed \cite{higherD, gaugingStates} with some first numerical applications on simple gauge theories. Here we will complement our study of the Schwinger model \cite{Buyens}. First, we will discuss the benefits of working on the gauge invariant subspace from the start, despite of Elitzurs theorem \cite{Elitzur} which states that the ground state on the full Hilbert space is automatically gauge invariant. Furthermore, we will show that we can correctly identify UV-divergent behaviour in our lattice computations. Specifically, we will study the UV-divergent chiral condensate, comparing our results with the analytic predictions \cite{Adam2, Durr} and previous numerical simulations \cite{BanulsProc}. \\
\\ \noindent\textit{ \bf Schwinger Hamiltonian.} The Schwinger model is (1+1)-d QED with one flavor. Its Lagrangian density reads $\mathcal{L} = \bar{\psi}\left(\gamma^\mu(i\partial_\mu+g A_\mu) - m\right) \psi - \frac{1}{4}F_{\mu\nu}F^{\mu\nu}\,.$
In the time-like axial gauge ($A_0=0$) one performs a Hamiltonian quantization, which can be turned in a lattice system by the Kogut-Susskind spatial discretization \cite{KogutSusskind}. In this formulation the two-component fermions are staggered: the first component lives on the even site, the second component on the odd site. After a Jordan-Wigner transformation, one obtains a spin-system coupled to a gauge-system:
\begin{equation}\label{equationH} H= \frac{g}{2\sqrt{x}}\Biggl( \sum_{n \in \mathbb{Z}} {L}^2(n) + \frac{\mu}{2} \sum_{n \in \mathbb{Z}}(-1)^n(\sigma_z(n) + (-1)^{n}) + x \sum_{n \in \mathbb{Z}}(\sigma^+ (n)e^{i\theta(n)}\sigma^-(n + 1) + h.c.)\biggl).\end{equation}
Here we have introduced the parameters $x \equiv 1/(g^2a^2)$, $\mu \equiv 2\sqrt{x}m/g$, with $a$ the lattice spacing, $m$ the fermion mass and $g$ the coupling constant. From now on we will work in units $g = 1$. The total Hilbert space is a tensor product of the local Hilbert spaces on the sites and on the links. On every site lives a spin system with $\sigma_z(n)\ket{s}_n= s \ket{s}_n (s=\pm 1)$. $\sigma^{\pm}=(1/2)(\sigma_x\pm i \sigma_y)$ are the ladder operators. On the links, where the gauge fields live, we have the operators $\theta(n)=a g A_1(na/2)$ and their conjugate momenta $L(n)$ ($[\theta(n),L(n')]=i\delta_{n,n'}$), which correspond to the electric field, $gL(n)=E(na/2)$. In a compact formulation $L(n)$ will have integer charge eigenvalues $p \in \mathbb{Z}$ and $e^{i\theta(n)}$ and $e^{-i\theta(n)}$ correspond to the ladders operators. So on link $n$ the local Hilbert space is spanned by the corresponding eigenkets $\ket{p}_n$ with $L(n)\ket{p}_n = p \ket{p}_n$ and $e^{\pm i\theta(n)}\ket{p}_n = \ket{p\pm 1}_n$.

The key-feature of the Hamiltonian (\ref{equationH}) is the gauge symmetry generated by $G(n) = L(n) - L(n-1) - (\sigma_z(n) + (-1)^n)/2$, which in the continuum limit corresponds to Gauss' law: $\partial_z E = gj^0$. All physical states $\ket{\Phi}$ then have to satisfy $G(n)\ket{\Phi} = 0, \forall n \in \mathbb{Z}$. Furthermore, the Hamiltonian is invariant under translations $T$, accompanied by a charge conjugation $C$, ($L \rightarrow - L$, $\sigma_z \rightarrow - \sigma_z$). This $CT$ symmetry is not spontaneously broken for all values of $m$ \cite{Coleman}.\\
\\ \noindent\textit{\bf Ground state. }In our approach we work in the thermodynamic limit $N \rightarrow \infty$ and block site $n$ and link $n$ into one MPS-site with local Hilbert space spanned by the kets $\ket{q}_n$ where ${q} = (s,p) (s = \pm 1, p \in \mathbb{Z})$. For the ground state we proposed in \cite{Buyens} the ansatz
\begin{equation}\label{CTMPS}\ket{\Psi(A)} = \sum_{q_n} v_L^\dagger \left(\prod_{n \in \mathbb{Z}}A^{q_n}\right) v_R \ket{\bm{q}^c}, \ket{\bm{q}^c} = \ket{\{(-1)^n q_{n}\}_{n \in \mathbb{Z}}},A^{q} \in \mathbb{C}^{D \times D},v_L, v_R \in \mathbb{C}^D.\end{equation}
One observes immediately that (\ref{CTMPS}) is manifestly $CT$ invariant. We also imposed the constraint
\begin{equation}\label{eqGIA} [A^{s,p}]_{(q,\alpha_q);(r,\beta_r)} =  \delta_{p, q + (s+1)/2} \delta_{r,-p}[\ta^{s,p}]_{\alpha_q, \beta_r}; q,r\in \mathbb{Z},\alpha_q,\beta_q = 1 \ldots D_q.\end{equation}
on the tensors, which makes our state gauge invariant: $G(n) \ket{\Psi(A)} = 0$. The variational freedom of this state thus lies within the matrices $\ta^{s,p}$. An approximation for the ground state was then obtained by doing imaginary time evolution ($\tau = it$) of the Schr\"odinger equation (SE), $i\partial_t \ket{\Psi(A)} = H\ket{\Psi(A)}$ using the time-dependent variational principle (TDVP), see \cite{Buyens,TDVP} for details. \\
\\ \noindent \textit{ \bf Gauge variant versus gauge invariant MPS.} By Elitzurs theorem \cite{Elitzur}, which states that a gauge symmetry cannot be spontaneously broken, one could argue that it is not necessary to impose the condition (\ref{eqGIA}) for a variational calculation of the ground state. However, there will typically be many more non-physical (gauge variant) low-energy excitations in the full Hilbert space, and one would therefore expect a slower convergence rate for variational calculations that do not impose gauge invariance. Let us now examine this issue explicitly for the Schwinger model. To this end we do a comparative study where we approximate the ground state with a MPS (\ref{CTMPS}), with and without imposing gauge invariance (\ref{eqGIA}). We take the parameters $m/g = 0.25$, $x = 100$ and do the simulations for $D = 29$ and $D = 40$. As explained in \cite{Buyens}, for the gauge invariant ansatz we have to distribute the variational freedom wisely among the charge sectors $D_q$  ($D = \sum_q D_q$) according to the Schmidt values. We truncate the charges $p$ on the links, $\vert p \vert \leq p_{max} = 2$ for $D = 29$ and $\vert p \vert \leq p_{max} = 3$ for $D = 40$. In all cases we used TDVP to find the ground state and stopped the algorithm when the norm of the gradient was below $10^{-6}$. In the second and fourth column of table \ref{table0} we display the simulations where we did not impose gauge invariance and in the third and fifth column the simulations where the states were manifestly gauge invariant. For reference, in the last column we display the $D=249$ result that was obtained in \cite{Buyens}. One immediately observes that the number of required steps is much larger in the gauge variant case. Furthermore, as the local dimension of the Hilbert space is larger, one TDVP iteration also takes more time in the gauge variant case. This leads to a huge difference in the total time: the gauge invariant simulations converged in a few minutes while the gauge variant simulations took a few hours. We can also explicitly verify Elitzurs theorem, by looking at the variance $\langle G^2 \rangle = \langle G(n)^2 \rangle, \forall n,$ of the gauge transformation generators for our ground state approximations on the full Hilbert space. As $\langle G^2 \rangle\approx 0$ we indeed converge to the gauge invariant ground state, which is also confirmed by the agreement of the ground state energy per site $e$ with the gauge invariant simulations.
\begin{table}
\begin{tabular}{|r|c|c|c|c|c|}
\hline
& without GI & with GI & without GI & with GI & with GI \cite{Buyens} \\
\hline
$p_{max}$ &$2$ & $2$& $3$ & $3$ & $3$ \\
$D$ & 29 & [2 6 9 8 4] & 40 & [2 3 7 11 10 4 2] &  [5 20 48 70 62 34 10] \\
 steps & 9645 & 278& 12417 &561 & \\
time &3 h 30 min  & 2 min & 6 h 27min & 5 min &\\
$\langle G^2 \rangle $ & $3 \times 10^{-9}$ & $0$ &$3 \times 10^{-9}$ & $0$ &$0$\\
$e$ & -3.048961 &--3.048961 &-3.048961 &-3.048961 & -3.048961 \\
 $E_{1,v}$& $1.04252 \{ 10 \}$&1.04254&1.04194 $\{ 14 \}$ &1.04209 & 1.04207 \\
 $E_{2,v} $& 2.455 $\{ 37 \}$ &2.455&2.385 $\{ 59 \}$ &2.386 & 2.357 \\
 $E_{1,s}$ &$1.7719 \{20 \}$ &1.7719&1.7559 $\{ 31 \}$&1.7565  & 1.7516 \\
\hline
\end{tabular}
\caption{Results of computations with and without imposing gauge invariance (GI). ($x = 100, m/g = 0.25$)}
\label{table0}
\end{table}

\begin{figure}
\begin{tabular}{rr}
\includegraphics[width=70mm]{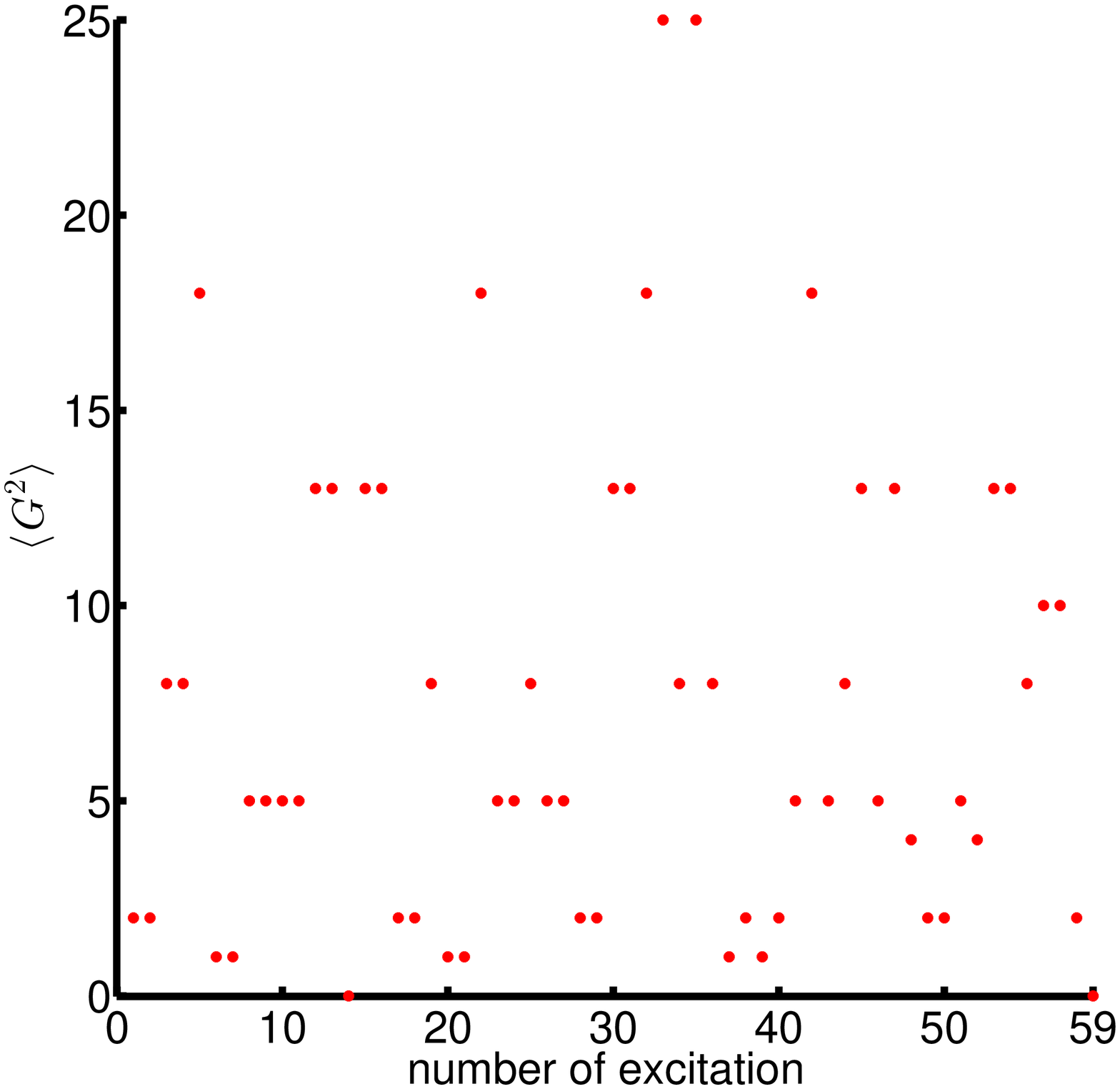} & \includegraphics[width=70mm]{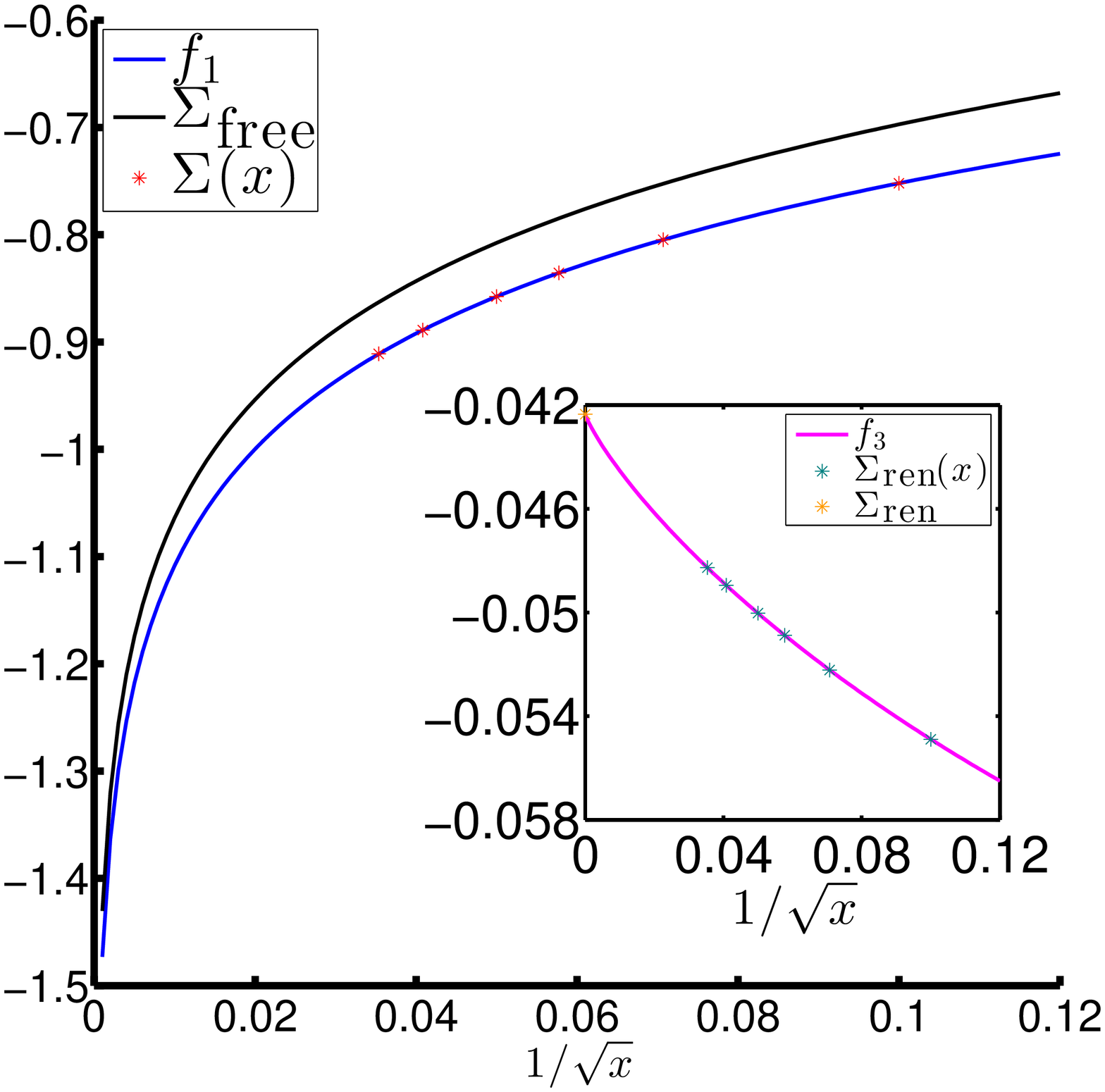}
\end{tabular}
\caption{Left (a): $m/g = 0.25, x = 100, D = 40:$ one-particle excitations with $CT = -1$, ranked according to increasing energy. Only those with $\langle G^2 \rangle = 0$ are gauge invariant. In this case, only the 14th and 59th excitations are physical and correspond to $E_{1,v}$ and $E_{2,v}$.  Right (b): $m/g = 0.5$: continuum extrapolation of the chiral condensate. Optimal fit $f_1(x) = A_1 + B_1\frac{\log(x)}{\sqrt{x}} + \Bigl( -\frac{m}{2g\pi} - C_1\Bigl)\log(x) + D_1\frac{1}{\sqrt{x}}$ (blue line) trough the data points $\Sigma(x)$ (red stars) for the five largest $x$. The divergence is removed by subtracting $\Sigma_{free}$ (black line). Inset: Optimal fit of  $f_3(x) = A_3 + B_3\frac{\log(x)}{\sqrt{x}} +  C_3\frac{1}{\sqrt{x}} + D_3\frac{1}{x}$ (magenta line) trough the data points $\Sigma_{ren}(x)$ (green stars) for the five largest $x$. The continuum value $\Sigma_{ren}$ is estimated as the intersection with the $y$-axis (orange star). }
\label{fig1}
\end{figure}

We have also examined the low-energy states which are computed with the same MPS-ansatz as in \cite{Buyens}, but now again with and without imposing gauge invariance. In \cite{Buyens} we found three stable one-particle excitations: two with $CT = -1$ and mass $E_{1,v}, E_{2,v}$ and one with $CT = 1$ and mass $E_{1,s}$. As expected, on the full Hilbert space we find many more non-physical excitations with $\langle G^2 \rangle \neq 0$. As illustrated in figure \ref{fig1} (a) we can identify the physical states by calculating $\langle G^2 \rangle$. The ranking (in increasing energy) where the physical excitations appear in the list of all excitations per sector ($CT=\pm 1$) is indicated in table \ref{table0} with curly brackets $\{\ldots \}$. There are indeed many gauge variant states lying between the ground state and low lying gauge invariant states. Moreover, the number of obtained gauge variant states increases with the bond dimension and we suspect that the Hamiltonian (\ref{equationH}) is gapless on the full Hilbert space.\\

\noindent\emph{\bf Chiral condensate.} For $m\neq 0$ the chiral condensate $\Sigma = \langle \bar{\psi} \psi\rangle$ is UV-divergent ($x\rightarrow \infty$), and it has been argued that also at the non-perturbative level, this divergence originates solely from the free ($g=0$) theory, leading to a logarithmic divergence, which is linear in $m$ \cite{ Adam2, Durr}. We now calculate the value of the chiral condensate with our MPS simulations and show that the scaling for large $x$ does indeed show the predicted UV-behavior. This problem was also studied in \cite{BanulsProc} with MPS-simulations for finite volumes. This allows us to compare results for the UV-regulated chiral condensate.
 
On the lattice $\Sigma = \langle \bar{\psi} \psi \rangle$ reduces to $\Sigma(x) = \frac{\sqrt{x}}{\vert \mathbb{Z}\vert }\sum_{n \in \mathbb{Z}}(-1)^n \Bigl\langle \frac{\sigma_z(n) + 1}{2}\Bigl\rangle$, which is easily computed from our MPS approximation (\ref{CTMPS}) for the ground state \cite{HaegemanMPSB}. For the Hamiltonian (\ref{equationH}), the free chiral condensate $\Sigma_{free}(x)$ can be computed exactly \cite{BanulsProc}:
\begin{equation}\label{SigmaFree} \Sigma_{free}(x) = -\frac{m}{\pi g}\frac{1}{\sqrt{1 +  \frac{m^2}{g^2 x}}}K\Biggl(\frac{1}{1 +  \frac{m^2}{g^2 x}}\Biggl)\end{equation}
where $K(z)$ is the complete elliptic integral of the first kind. As $x \rightarrow \infty$, we indeed have up to finite terms $\Sigma_{free}(x) \rightarrow -m/(2g\pi)\log(x)$. We now verify that this is the only UV-divergence for all values of $m/g$. Thereto we compute $\Sigma(x)$ for $x = 100, 200, 300, 400, 600, 800$. We find that $f_1(x) = A_1 + B_1\frac{\log(x)}{\sqrt{x}} + \Bigl( -\frac{m}{2g\pi} - C_1\Bigl)\log(x) + D_1\frac{1}{\sqrt{x}}$ results in a good fit to the data $\Sigma(x)$, see figure \ref{fig1} (b). Our estimate of $C_1$ is obtained by $i)$ fitting $\Sigma(x)$ to $f_1(x)$ for the five largest $x$, $ii)$ fitting all the data to $f_1(x)$, $iii)$ fitting all the data to $f_2(x) =  A_2 + B_2\frac{\log(x)}{\sqrt{x}} + \Bigl( -\frac{m}{2g\pi} - C_1\Bigl)\log(x) + D_2\frac{1}{\sqrt{x}} + E_2 \frac{1}{x}$. The displayed value of $C_1$ in the second column of table \ref{table1} is the one with the largest magnitude of the fits $i), ii)$ and $iii)$. We observe that $C_1\approx0$, consistent with the claim \cite{Adam2} that the full non-perturbative $UV$-divergence can indeed be traced back completely to the free chiral condensate (\ref{SigmaFree}).

To compare our results with \cite{BanulsProc}, we renormalize the chiral condensate by subtracting $\Sigma_{free}(x)$ from $\Sigma(x)$. As in \cite{BanulsProc} we fit $f_3(x) = A_3 + B_3\frac{\log(x)}{\sqrt{x}} +  C_3\frac{1}{\sqrt{x}} + D_3\frac{1}{x}$ to the renormalized chiral condensate $\Sigma_{ren}(x) = \Sigma(x) - \Sigma_{free}(x)$. Our estimate for $\Sigma_{ren}$ is the $A_3$ obtained by a fit through the largest five $x$-values (see figure \ref{fig1} (b), inset). The error on this value is estimated as the maximum of the difference with the $A_3$'s we would obtain if we fitted all data to $f_3(x)$ and to $f_4(x)  = A_3 + B_4\frac{\log(x)}{\sqrt{x}} +  C_4\frac{1}{\sqrt{x}} + D_4\frac{1}{x} + E_4\frac{1}{x^{3/2}}$. This error dominates the error due to the truncation of the bond dimension. The results can be found in the third column of table \ref{table1}. We see that our results agree very well with \cite{BanulsProc} and with the exact strong coupling ($m/g=0$) result: $\Sigma_0=-e^\gamma/(2\pi^{3/2})\approx-0.1599288$ . 
\begin{table}
\begin{center}
\begin{tabular}{|c||c|c|c|c|c| }
\hline
$m/g$  & $C_1$ &$\Sigma_{ren}$ & $\Sigma_{ren} $\cite{BanulsProc} & exact   \\
\hline
0       & $3 \times 10^{-6}$ & -0.159928 (1)  &-0.159930 (8) & -0.1599288  \\
0.125&$3\times 10^{-5}$ &-0.092019 (2)  &-0.092019 (4)  & -\\
0.25 & $4\times 10^{-5}$&-0.066647 (4) &-0.066660 (11)&- \\
0.5 &    $1\times 10^{-4}$    &-0.042349 (2)& -0.042383 (22) &-\\
0.75 &       $2 \times 10^{-4}$ &-0.03062 (3) & - & -\\
1 & $3 \times 10^{-4}$  &-0.023851 (8)  & - & - \\
2 &$1\times 10^{-3}$      &-0.012463 (9)  &-  &- \\
\hline
\end{tabular}
\end{center}
\caption{Results for chiral condensate.}
\label{table1}
\end{table}
\\
\\ \noindent\textit{\bf Conclusions.} In this proceeding we continued the exploration of the Schwinger model as a testbed for MPS simulations of Hamiltonian lattice gauge theories. In the time-like axial gauge, the theory is perfectly well defined on the full (positive norm) Hilbert space, including many gauge variant states, but with a gauge invariant ground state due to Elitzurs theorem. Nevertheless, we have explicitly demonstrated that the computation time for simulations on this full Hilbert space scales unfavorably and it is of paramount importance to exploit the gauge invariance, by working with gauge invariant trial MPS states. We have also calculated the chiral condensate, thereby recovering the proper UV-divergent behavior in the continuum limit.

It is clear by now that the MPS framework offers a reliable method for the simulation of 1+1 dimensional lattice gauge theories, even in the continuum limit. It will certainly be interesting to further explore (also non-abelian) gauge theories with MPS, especially in those regimes (real-time dynamics and finite fermionic densities) that are inaccessible to the lattice Monte-Carlo simulations. Of course a more important and more challenging goal is to extend the TNS approach to simulations of higher dimensional gauge theories. The higher dimensional generalization of MPS go by the name of projected entangled pair states (PEPS) \cite{TNS}. The present algorithms only allow PEPS calculations with relatively small bond dimensions, and the development of the TNS framework towards a quantitative method for gauge theories like QCD will most probably require better algorithms. Nevertheless, one can use already PEPS with small bond dimensions to study certain model Hamiltonians. This approach was pursued recently to explore the phase diagram of $Z_2$ gauge theory \cite{gaugingStates} in $d=2+1$; and it should certainly be interesting to generalize this approach to other gauge theories, and even include fermions. \\

\noindent\textit{Acknowledgements. } This work is supported by an Odysseus grant from the FWO, a PhD-grant from the FWO (B.B.), a postdoc-grant from the FWO (J.H.), the FWF grants FoQuS and Vicom, the ERC grant QUERG and the EU grant SIQS.

\end{document}